\documentclass[a4paper,12pt]{article}
\usepackage{amssymb} 
\usepackage{amsmath,amsfonts,epsfig,subfigure}

\begin{document}
\title{Stoneley waves and interface stability \\of Bell materials in 
compression; \\  Comparison with rubber.}

\author{Michel Destrade}
\date{2005}
\maketitle  

\bigskip

\begin{abstract}
\noindent
Two semi-infinite bodies made of prestressed, homogeneous, 
Bell-constrained, hyperelastic materials are perfectly bonded along a 
plane interface.
The half-spaces have been subjected to finite pure homogeneous 
predeformations, with distinct stretch ratios but common principal 
axes, and such that the interface is a common principal plane of 
strain.
Constant loads are applied at infinity  to maintain the
deformations and the influence of these loads on the propagation of
small-amplitude interface (Stoneley) waves is examined.
In particular, the secular equation is found and 
necessary and sufficient conditions to be satisfied by the stretch 
ratios to ensure the existence of such waves are given.
As the loads vary, the Stoneley wave speed varies accordingly:
the upper bound is the `limiting speed' (given explicitly), beyond
which the wave amplitude cannot decay away from the interface;
the lower bound is zero, where the interface might become unstable.
The treatment parallels the one followed for the incompressible case
and the differences due to the Bell constraint are highlighted.
Finally, the analysis is specialized to specific strain energy 
densities and to the case where the bimaterial is uniformly deformed 
(that is when the stretch ratios for the upper half-space are equal to 
those for the lower half-space.)
Numerical results are given for `simple hyperelastic Bell' materials 
and for `Bell's empirical model' materials, and compared to the 
results for neo-Hookean incompressible materials.

\end{abstract}

\newpage

\section{INTRODUCTION}

This paper is a prolongation of a series of articles concerned with 
the propagation of small-amplitude waves in deformed Bell materials.
Beatty and Hayes \cite{BeHa95} wrote the incremental equations of 
motion for a hyperelastic body maintained in a state of static finite
homogeneous deformation and subject to the constraint of Bell 
\cite{BeHa92a}: tr $\mathbf{V} = 3$, where $\mathbf{V}$ is the left
stretch tensor associated with the finite deformation.
They studied plane \cite{BeHa95} as well as torsional \cite{BeHa95b}
waves.
Then this author \cite{Dest02b} found the secular equation for surface 
(Rayleigh) waves propagating on a semi-infinite deformed body made 
of Bell-constrained material.
Next \cite{Dest03} the surface stability of such a half-space under
compression was analyzed and compared with that of incompressible 
rubber.
Now the propagation of interface (Stoneley) waves is considered for
two rigidly bonded semi-infinite bodies made of distinct Bell 
materials.
Each half-space was subjected to a finite homogeneous predeformation 
and the principal axes of each deformation are assumed to coincide at 
the interface, which is a common principal plane.
Moreover, the interfacial wave is assumed to propagate in a principal
direction.
These common \cite{DoOg91,Chad95} restrictions aside, a general 
necessary and sufficient condition of existence of a Stoneley 
wave is established, together with the corresponding secular equation.

These general results are obtained in Section 3, after the basic
equations of the problem have been written in Section 2.
For purposes of comparison and contrast, these latter equations are
written in a manner which is similar to those governing the
propagation of interfacial waves in deformed hyperelastic materials
subject to the constraint of incompressibility, det $\mathbf{V} = 1$.
Then in Section 4, the analysis is specialized to the case where the 
infinite body made of the two bonded half-spaces is uniformly 
prestrained, that is when the principal stretch ratios of strain are
the same for the upper half-space and for the lower half-space.
For ``simple hyperelastic Bell'' materials, the influence of the 
initial deformation on the wave speed is highlighted.
As the half-spaces are more and more compressed, this speed tends to
zero and the secular equation gives the ``neutral equation''.
The critical stretch ratios, at which the neutral equation is 
reached, are obtained.
In the case of a plane prestrain, these critical 
stretches, and the corresponding ones for ``Bell's empirical model'', 
are compared with the critical stretches obtained by Biot 
\cite{Biot63} for the interfacial stability of (incompressible) 
neo-Hookean materials, often used to model rubber.
The conclusion is that, as far as interfacial stability is concerned,
rubber may be compressed more than Bell's empirical materials,
but not as much as Bell simple hyperelastic materials, before the 
critical stretches are reached. 

\section{PRELIMINARIES}
%
\subsection{Deformed Bell half-spaces}
%
Consider two semi-infinite bodies made of different homogeneous 
Bell-con\-strai\-ned hyperelastic materials, 
separated by plane interface $x_2=0$. 
Initially, each body is at rest in a reference configuration, denoted 
by $\mathcal{B}_0$ for the lower half-space $x_2 \ge 0$, and
by $\mathcal{B}^*_0$ for the upper half-space $x_2 \le 0$.
The constitutive equation for the lower half-space is \cite{BeHa92a}
\begin{equation} \label{constitutive} 
\mathbf{T}= p \mathbf{V} + \omega_0 \mathbf{1}
  +  \omega_2 \mathbf{V}^2,
\end{equation}
expressing the Cauchy stress tensor $\mathbf{T}$ in terms of the 
left stretch tensor $\mathbf{V}$.
Here $p$ is an undetermined scalar, to be found from the equations of
motion or of equilibrium, and from the boundary conditions.
The constitutive parameters $ \omega_0$ and $ \omega_2$ are defined 
in terms of the derivatives of the strain energy density 
$\Sigma = \Sigma(i_2,i_3)$ of the hyperelastic body with respect to 
the principal invariants 
$i_2 = [ (\text{tr } \mathbf{V})^2 - \text{tr } (\mathbf{V}^2)]/2$, 
$i_3 = \text{det } \mathbf{V}$, as
\begin{equation} \label{A-inequalities}
\omega_0 = \partial \Sigma / \partial i_3 \le 0, 
\quad
\omega_2 = - i_3^{-1} \partial \Sigma / \partial i_2 > 0.
\end{equation}
The inequalities above are called the `Beatty-Hayes $A$-inequalities' 
\cite{BeHa92a}.
Also, the Bell constraint is satisfied at all times:
\begin{equation} \label{Bell}
i_1 \equiv \text{tr } \mathbf{V} =3.
\end{equation}
Similar (starred) equalities and inequalities apply for the upper 
half-space $x_2 \le 0$.

Next, consider that each half-space is subject to a finite static pure 
homogeneous static deformation 
$\mathcal{B}_0 \rightarrow \mathcal{B}_1$ and
$\mathcal{B}^*_0 \rightarrow \mathcal{B}^*_1$
by the application of suitable loads.
The principal axes of these deformations along $x_2$, and $x_1$ 
and $x_3$, two orthogonal directions in the $x_2=0$ plane.
Hence the finite deformations are described in the coordinate system
of the principal axes by
\begin{equation}
x_\Gamma = \lambda_\Gamma X_\Gamma 
\quad (x_2 \ge 0), \quad \text{and} \quad
x_\Gamma = \lambda^*_\Gamma X_\Gamma 
\quad (x_2 \le 0), \quad (\Gamma = 1,2,3; \text{no sum}),
\end{equation}
where the $\lambda_\Gamma$,  $\lambda^*_\Gamma$ are the
principal stretch ratios for each half-space.
For the lower half-space we have
\begin{align} \label{invariants-static}
& \mathbf{V} = \text{Diag}(\lambda_1, \lambda_2, \lambda_3),
\notag \\
& \lambda_1 + \lambda_2 + \lambda_3 = 3,
\quad
i_2 = \lambda_1 \lambda_2
   + \lambda_2 \lambda_3 + \lambda_3 \lambda_1,
\quad
i_3 = \lambda_1 \lambda_2 \lambda_3,
\end{align}
and the following constant Cauchy stress tensor 
\begin{equation}  \label{T}
T_{\Gamma \Gamma} 
 = p_o \lambda_\Gamma + \omega_0
      + \omega_2 \lambda_\Gamma^2
\quad (\Gamma = 1,2,3; \text{no sum}), 
\quad
T_{ij} = 0
\quad
(i \ne j),
\end{equation}
where $p_o$ is constant, and $\omega_0$, $\omega_2$ are given by  
\eqref{A-inequalities} and valuated at $i_2$, $i_3$ given by
\eqref{invariants-static}$_{3,4}$, clearly satisfies the equilibrium 
equations $T_{ij,j}=0$.
So does $\mathbf{T}^*$, the equivalent starred version for 
the upper half-space.
Finally, the half-spaces are rigidly bounded. 
At the interface $x_2=0$, the continuity of the tractions imposes that
\begin{equation} \label{po}
p_o \lambda_2 + \omega_0   + \lambda_2^2 \omega_2
 = p_o^* \lambda^*_2 + \omega^*_0   + \lambda^{*2}_2 \omega^*_2.
\end{equation}
The half-spaces are maintained in these deformed states
$\mathcal{B}_1$, $\mathcal{B}^*_1$ 
by the application at infinity of the loads 
$P_\Gamma = - T_{\Gamma \Gamma}$, 
$P^*_\Gamma = -T^*_{\Gamma \Gamma}$.
It follows from  \eqref{po} that $P_2 = P^*_2$.
%
\subsection{Small-amplitude interface wave}
%

Now consider that a small amplitude wave travels with speed $v$
and a wave number $k$ at the interface, in the direction of $x_1$,
with attenuation away from $x_2=0$.
The following expressions describe this incremental motion 
$\mathcal{B}_1 \rightarrow \overline{\mathcal{B}}_1$, 
$\mathcal{B}^*_1 \rightarrow \overline{\mathcal{B}}^*_1$:
\begin{equation}  \label{xBar}
\overline{\mathbf{x}} = \mathbf{x} 
    + \varepsilon \Re \{ \mathbf{U}(kx_2) e^{i k (x_1 - vt)} \},
\quad \text{and} \quad
\overline{\mathbf{x}} = \mathbf{x}
    + \varepsilon \Re \{ \mathbf{U}^*(kx_2) e^{i k (x_1 - vt)} \},
\end{equation}
for $x_2 \ge 0$ and for $x_2 \le 0$, respectively.
Here $\varepsilon$ is small enough to allow 
linearization and the amplitudes $\mathbf{U}$ and $\mathbf{U}^*$
are of the form $\mathbf{U}= [U_1, U_2, 0]^{\mathrm{T}}$, 
$\mathbf{U}^*= [U^*_1, U^*_2, 0]^{\mathrm{T}}$.
The remainder of this paper is devoted to the study of these 
interfacial (Stoneley) waves.
Dowaikh and Ogden \cite{DoOg91} and Chadwick \cite{Chad95} conducted
similar studies for prestrained incompressible and compressible 
materials, respectively.
In both cases, the authors proved that the initial normal traction
$T_{22} = T^*_{22}$ played no role in the resolution and analysis of 
the problem.
It can be checked that such is also the case for deformed Bell 
materials and without loss of generality, we choose 
$T_{22} = T^*_{22}=0$, and thus $P_2 = P_2^*=0$ 
(of course, the traction $T_{22}$ might play a role in issues other 
than the derivation of the secular equation for Stoneley waves.)
Hence the following loads $P_1 = - T_{11}$, $P_3 = -T_{33}$ and 
$P^*_1$, $P^*_3$ are applied at $x_1=\pm \infty$ and $x_3=\pm \infty$
in order to sustain the finite homogeneous deformation,
\begin{align} \label{P1P3}
& P_\Gamma =  (\lambda_2 -  \lambda_\Gamma)
         (-\omega_0 + \lambda_\Gamma \lambda_2 \omega_2)/ \lambda_2,
\notag \\
& P^*_\Gamma =  (\lambda^*_2 -  \lambda^*_\Gamma) 
   (-\omega^*_0 +\lambda^*_\Gamma\lambda^*_2 \omega^*_2)/\lambda^*_2,
\notag \\
& (\Gamma =1,3).
\end{align}
Moreover, this assumption enables us to use previously established 
results for the incremental equations of motion.
Specifically, the equations are written as the following system of
first order differential equations for the components of the 
displacements $\mathbf{U}$, $\mathbf{U}^*$, and of the tractions 
$\mathbf{t}$, $\mathbf{t}^*$ on the planes $x_2=$ const.  
(see Ref.\cite{Dest02b} for details):
\begin{align} \label{quartic}
&  b_3 \lambda_2^2 U_1' + ib_3 \lambda_2^2 U_2 -  t_1 = 0, 
&& t_1' + i \lambda_1 \lambda_2^{-1} t_2
    - (\lambda_1 \lambda_2^{-1}C - \rho v^2) U_1 = 0,  \nonumber 
\\
&  U_2' + i \lambda_1 \lambda_2^{-1} U_1 = 0, 
&& t_2' + i  t_1 
        -[b_3(\lambda_1^2 -  \lambda_2^2) - \rho v^2 ] U_2= 0, 
\end{align}
where 
\begin{align} \label{coefficients}
& b_3 = \frac{-\omega_0 + \lambda_1 \lambda_2 \omega_2} 
        {\lambda_2(\lambda_1 + \lambda_2)},
\nonumber 
\\
& 
C = \lambda_1^{-1} \lambda_2 C_{11}
    + \lambda_1 \lambda_2^{-1} C_{22} -C_{12} - C_{21} 
       -2\omega_0 - (\lambda_1^2+\lambda_2^2)\omega_2,
\nonumber 
\\
& C_{ij}  = 
   2 \lambda^2_{i} \delta_{ij} \omega_2
    - \lambda_{j}^2 (\omega_{02} + \lambda_{i}^2 \omega_{22}) 
    + \lambda_1 \lambda_2  \lambda_3 (\omega_{03} + 
                          \lambda_{i}^2 \omega_{23}),
\end{align}
for $x_2 \ge 0$, and a starred version for $x_2 \le 0$.
Here $\rho$ ($\rho^*$) is the mass density of the lower (upper) 
half-space, and the derivatives $\omega_{0\Gamma}$, $\omega_{2\Gamma}$
($\Gamma = 2,3$) of the material parameters $\omega_0$, $\omega_2$ 
are taken with respect to $i_\Gamma$ and evaluated at $i_2$, $i_3$ 
given by \eqref{invariants-static}.
The half-spaces are rigidly bonded at $x_2=0$ so that the following 
boundary conditions apply,
\begin{equation}    \label{BC1}
\mathbf{U}(0) = \mathbf{U}^*(0), \quad 
\mathbf{t}(0) = \mathbf{t}^*(0),
\end{equation}
together with the requirement that the wave vanishes as 
$x_2 \rightarrow \pm \infty $. 

%
\section{GENERAL PREDEFORMATIONS AND STRAIN ENERGY FUNCTIONS}
%

Here we solve the problem at hand, that is we solve the equations of
motion \eqref{quartic} and then apply the boundary conditions 
\eqref{BC1} in order to derive the secular equation for the speed of
Stoneley waves in deformed Bell materials.
We adopt an approach reminiscent of that used by Dowaikh and 
Ogden \cite{DoOg91} for Stoneley waves in incompressible materials and
show how each constraint leads to different results.

\subsection{Secular equation; comparison with incompressible materials}

For small-amplitude waves of the form \eqref{xBar}$_1$,
the incremental constraint of incompressibility, 
$u_{1,1} + u_{2,2} + u_{3,3} = 0$, imposes that
\begin{equation}
i U_1 + U'_2 = 0,
\end{equation}
suggesting the introduction of the function $\hat{\varphi}$ defined by
\begin{equation} \label{Uphi}
U_1(k x_2) = i  \hat{\varphi}'(k x_2), 
\quad
U_2(k x_2) =   \hat{\varphi}(k x_2).
\end{equation}
 
In our context, the incremental constraint of Bell, 
$\lambda_1 u_{1,1} + \lambda_2 u_{2,2} + \lambda_3 u_{3,3} = 0$, 
imposes Eq.\eqref{quartic}$_3$, that is
\[  
i \lambda_1 \lambda_2^{-1} U_1 + U'_2 = 0,  
\]
suggesting the function $\varphi$ defined by
\begin{equation} \label{phiHat}
U_1(k x_2) = i  \varphi'(k \lambda_1 \lambda_2^{-1}  x_2), 
\quad
U_2(k x_2) =   \varphi(k \lambda_1 \lambda_2^{-1}  x_2).
\end{equation}
From \eqref{quartic}$_{4,1}$, the traction components are also 
expressed in terms of $\varphi$ and its derivatives:
\begin{equation} \label{tphi}
t_1 = i b_3 \lambda_2^2(\lambda_1 \lambda_2^{-1} \varphi'' 
       		+  \varphi), \quad
t_2 = - b_3 \lambda_1 \lambda_2 \varphi'''  
       		+  \lambda_1^{-1} \lambda_2(\lambda_1 \lambda_2^{-1} C 
        	- b_3 \lambda_1 \lambda_2 - \rho v^2) \varphi'.
\end{equation}
Finally, Eq.\eqref{quartic}$_2$ reads
\begin{equation} \label{eqWithPhi1st}
b_3 \lambda_1^2 \varphi'''' 
 - (\lambda_1 \lambda_2^{-1} C
     - 2b_3 \lambda_1 \lambda_2 - \rho v^2)\varphi''
	+ (b_3 \lambda_1^2 - \rho v^2)\varphi = 0.
\end{equation}
The same procedure is of course also valid for the upper half-space, 
with the introduction of a function $\varphi^*$.

Dowaikh and Ogden \cite{DoOg91} expressed the strain energy
density as a function $\hat{W}(\lambda_1, \lambda_2, \lambda_3)$ 
of the principal stretches of deformation, 
rather than as a function of the invariants 
$\hat{\Sigma}(i_1, i_2, i_3)$,
a practice also favored by Biot \cite{Biot65}.
With this choice, and the introduction of the function $\hat{\varphi}$
in \eqref{phiHat}, they proved that for \textit{incompressible 
materials}, the counterpart to \eqref{eqWithPhi1st} is 
\begin{equation} \label{eqWithPhiHat}
\hat{\gamma } \hat{\varphi}'''' 
 - (2 \hat{\beta } - \rho v^2)\hat{\varphi}''
	+ (\hat{\alpha }- \rho v^2)\hat{\varphi} = 0,
\end{equation}
where $ \hat{\alpha}$, $\hat{\beta}$, and $\hat{\gamma}$ are defined in
terms of $\hat{W}$ and its derivatives with respect to the $\lambda_i$
as
\begin{align}
& \hat{\alpha} \lambda_2^2 = \hat{\gamma} \lambda_1^2 
= (\lambda_1 \hat{W}_1 - \lambda_2 \hat{W}_2)
     \lambda_1^2 \lambda_2^2/(\lambda_1^2 - \lambda_2^2), 
\nonumber \\
& 2\hat{\beta} + 2 \hat{\gamma} =
  \lambda_1^2 \hat{W}_{11} + \lambda_2^2 \hat{W}_{22} 
    - 2 \lambda_1 \lambda_2  \hat{W}_{12} 
	+  2\lambda_2 \hat{W}_2.
\end{align} 
Also, for incompressible materials, the assumption of strong 
ellipticity for the equations of motion implies that
\begin{equation}
\hat{\alpha}>0, \quad
\hat{\gamma} >0, \quad
\hat{\beta}+ \sqrt{\hat{\alpha}\hat{\gamma}} >0.
\end{equation}

For \textit{Bell constrained materials}, introduce  by analogy 
$\alpha$, $\beta$, and $\gamma$,
defined by
\begin{equation}
 \alpha  = b_3 \lambda_1^2, 
\quad 
 \gamma = b_3 \lambda_2^2,  
\quad
 2 \beta + 2 \sqrt{\alpha \gamma} =
  \lambda_1 \lambda_2^{-1} C, 
\end{equation} 
or equivalently, by
\begin{align} \label{AlphaBetaGamma}
& \alpha \lambda_2^2 = \gamma \lambda_1^2 = 
    (W_1 - W_2) 
       \lambda_1^2 \lambda_2/[\lambda_3 (\lambda_1^2 - \lambda_2^2)], 
\nonumber \\
& 2\beta + 2 \sqrt{\alpha \gamma} = 
  (W_{11} + W_{22} - 2 W_{12}) \lambda_1/(\lambda_2 \lambda_3).
\end{align} 
Now  rewrite the equation of motion \eqref{eqWithPhi1st} as 
\begin{equation} \label{eqWithPhi2nd}
\alpha  \varphi''''  - (2 \beta  - \rho v^2)\varphi''
	+ (\alpha -  \rho v^2)\varphi = 0. 
\end{equation}
Using the strong ellipticity condition for the equations of motion in 
deformed Bell materials (see Appendix) together with 
Eqs.\eqref{A-inequalities} and \eqref{coefficients}, 
we also find the following inequalities,
\begin{equation} \label{SEBell}
\alpha >0, \quad
\gamma >0, \quad
\beta + \alpha  >0.
\end{equation}
The contrast between the effects of incompressibility and of the Bell 
constraint is striking
(compare for instance the first term in \eqref{eqWithPhiHat} and in 
\eqref{eqWithPhi2nd}.)
Now Equation \eqref{eqWithPhi2nd} and its starred version for the
upper half-space are solved. 

Because Stoneley waves vanish with distance from the interface,
the solutions $\varphi$ and $\varphi^*$ must be of the form
\begin{equation} \label{PhiPhi*}
\varphi(z) = A e^{-s_1z} + B e^{-s_2z} \quad (z \ge 0), \quad
\varphi^*(z) = A^* e^{s_1^*z} + B^* e^{s_2^*z} \quad (z \le 0), 
\end{equation}
where $s_1 \ne s_2$, $\Re(s_i)>0$, $s^*_1 \ne s^*_2$, $\Re(s^*_i)>0$, 
for some constants $A$, $B$, $A^*$, $B^*$.
Explicitly, the $s_i$, $s^*_i$ are roots of the biquadratics
\begin{align} \label{biquadratic}
& \alpha s^4 - (2\beta - \rho v^2)s^2 + \alpha - \rho v^2 = 0,
& \alpha^* s^{*4} - (2\beta^* - \rho^* v^2)s^{*2} 
      + \alpha^* - \rho^* v^2 = 0,  \nonumber \\
& s_1^2 + s_2^2 = (2\beta - \rho v^2)/\alpha, 
& s_1^{*2} + s_2^{*2} = (2\beta^* - \rho^* v^2)/\alpha^*,  \nonumber \\
& s_1^2 s_2^2 = (\alpha - \rho v^2)/\alpha, 
& s_1^{*2} s_2^{*2} = (\alpha^* - \rho^* v^2)/\alpha^*.  
\end{align}
The roots $s_1^2$ and $s_2^2$ of the real quadratic 
\eqref{biquadratic}$_1$ are either both real or both complex;
if they are real, they are non-negative, otherwise $s_1$ and $s_2$ 
are purely imaginary in contradiction with the decaying requirement
$\Re(s_i)>0$;
if they are complex, they are conjugate, because the coefficients of 
\eqref{biquadratic}$_1$ are real.
In both cases, $s_1^2 s_2^2 \ge 0$ 
and similarly, $s_1^{*2} s_2^{*2} \ge 0$, so that 
\begin{equation} \label{intervalForV}
0 \le v \le \text{min}\{v_L,v^*_L\}, \quad \text{where} 
\quad v_L = \sqrt{\alpha/\rho}, \; v^*_L = \sqrt{\alpha^*/\rho^*}.
\end{equation} 
In fact, the interval of possible values for $v$ for which the wave 
decays away from the interface (the \textit{subsonic  interval})
might be smaller than the interval defined above.
This point is clarified in \S \ref{The-subsonic-interval}.
For the time being, we derive an explicit expression for the secular
equation.
Equations \eqref{BC1} express the continuity of the displacement and
 traction incremental amplitudes at the interface $x_2=0$.
Using the expressions \eqref{Uphi}, \eqref{tphi} of $U_i$, $U_i^*$,
$t_i$, $t^*_i$, in terms of $\varphi$, $\varphi^*$ and their 
derivatives, they lead to
\begin{align}
& \varphi(0)=\varphi^*(0), \quad \varphi'(0)=\varphi^{*'}(0), 
\nonumber \\
& \gamma[\sqrt{\frac{\gamma}{\alpha}}\varphi''(0) +\varphi(0)] =
 \gamma^*[\sqrt{\frac{\gamma^*}{\alpha^*}}\varphi^{*''}(0)
                                               +  \varphi^*(0)],
\\
&-\sqrt{\alpha\gamma}\varphi'''(0) 
 + \sqrt{\frac{\gamma}{\alpha}}
   (2\beta+ \sqrt{\alpha\gamma}-\rho v^2)\varphi'(0) = 
\nonumber \\
& \phantom{-\sqrt{\alpha\gamma}\varphi'''(0) 
 + \sqrt{\frac{\gamma}{\alpha}}}
-\sqrt{\alpha^*\gamma^*}\varphi^{*'''}(0) 
 + \sqrt{\frac{\gamma^*}{\alpha^*}}
   (2\beta^* + \sqrt{\alpha^*\gamma^*}-\rho v^2)\varphi^{*'}(0).
\nonumber
\end{align}
Using \eqref{biquadratic}$_{3,4}$ and \eqref{PhiPhi*}, we rewrite this
system as four homogeneous linear equations for the unknowns $A$, $B$,
$A^*$, $B^*$:
\begin{align}
& A+B = A^*+B^*, \nonumber \\
& s_1 A + s_2 B =  -s^*_1 A^* - s^*_2 B^*, 
\nonumber \\
&(\sqrt{ \alpha \gamma}s_1^2 +\gamma)A
 + (\sqrt{\alpha \gamma}s_2^2 +\gamma)B =\nonumber\\
& \quad \quad   \quad \quad  
(\sqrt{\alpha^* \gamma^*}s_1^{*2} +\gamma^*)A^*
 + (\sqrt{\alpha^* \gamma^*}s_2^{*2} +\gamma^*)B^*,
\nonumber\\
& s_1(\sqrt{\alpha\gamma}s_2^2 + \gamma)A +
   s_2(\sqrt{\alpha\gamma}s_1^2 + \gamma)B = \nonumber\\
& \quad \quad   \quad \quad  
-s^*_1(\sqrt{\alpha^*\gamma^*}s_2^{*2} + \gamma^*)A^* 
   -s^*_2(\sqrt{\alpha^*\gamma^*}s_1^{*2} + \gamma^*)B^*.
\end{align}
The vanishing of the corresponding determinant yields the secular
equation.
Dropping the factor $(s_1-s_2)(s^*_1-s^*_2)$ and using the quantities
$\eta$, $r$ (and $\eta^*$, $r^*$) defined as follows,
\begin{equation} \label{r-eta}
\eta= \sqrt{\frac{\alpha-\rho v^2}{\alpha}}, \quad
r = \sqrt{ (\eta+1)^2 + 2\frac{\beta - \alpha}{\alpha}},
\end{equation}
the \textit{secular equation} is written compactly as
\begin{align} \label{secular}
i(v) & \equiv  
\gamma^2[\frac{\alpha}{\gamma}\eta r^2 
   - (\sqrt{\frac{\alpha}{\gamma}}\eta -1)^2] + 
\gamma^{*2}[\frac{\alpha^*}{\gamma^*}\eta^* r^{*2} 
   - (\sqrt{\frac{\alpha^*}{\gamma^*}}\eta^* -1)^2]  \nonumber
\\
 & \quad + \gamma \gamma^*
 [\sqrt{\frac{\alpha \alpha^*}{\gamma \gamma^*}}(\eta + \eta^*) r r^* 
   +  2(\sqrt{\frac{\alpha}{\gamma}}\eta -1)
          (\sqrt{\frac{\alpha^*}{\gamma^*}}\eta^* -1)] \\
& =  0. \nonumber
\end{align}
The quantities $\eta$, $\eta^*$ are real by \eqref{intervalForV}
and so are $r$, $r^*$, at least as long as the decaying condition 
is satisfied.
We now elaborate on this last point.

\subsection{The subsonic interval}
\label{The-subsonic-interval}

From \eqref{biquadratic} and the definition \eqref{r-eta} of $r$,
expressions for the squared sum and difference of the roots $s_i$ 
follow:
\begin{equation}
(s_1+s_2)^2 = r^2, \quad (s_1 - s_2)^2 = s^2, 
\end{equation}
where 
\begin{equation}
s^2 = \frac{2\beta - \rho v^2}{\alpha}
        - 2\sqrt{\frac{\alpha - \rho v^2}{\alpha}} 
 = (\eta - 1)^2 + 2\frac{\beta - \alpha}{\alpha}
 = r^2 - 4\eta.
\end{equation}
Hence the requirement of exponential decay $\Re(s_i)>0$ is equivalent
to 
\begin{equation} \label{decay}
\Re(r \pm s) > 0.
\end{equation}
Now we span the \textit{subsonic} range, that is the interval $I$
(say) of possible values for $v$ such that \eqref{decay} is satisfied.
This interval is $I = [0, \hat{v}]$ where $\hat{v}$ is called the
\textit{limiting speed}.

Starting at $v=0$, we have, using the strong ellipticity condition
\eqref{SEBell}$_3$,
\begin{equation}
v=0, \quad \eta=1, \quad r = \sqrt{2\frac{\beta+\alpha}{\alpha}}>0,
\quad r^2 - s^2 = 4 >0,
\end{equation}
so that \eqref{decay} is satisfied.

Now, if $2\beta-\alpha>0$, then we may increase $v$ from $0$ to 
$v_L=\sqrt{\alpha/\rho}$, where
\begin{equation}
v=v_L, \quad \eta=0, \quad r = \sqrt{\frac{2\beta-\alpha}{\alpha}}>0,
\quad r^2 - s^2 = 0,
\end{equation}
and $\eta$, $r$, $r^2 - s^2$, decrease monotonically with $v$ 
but remain non-negative real numbers, so that \eqref{decay} is 
satisfied over the interval $[0,v_L]$.
Clearly in this situation, the limiting speed is 
$\hat{v}=v_L = \sqrt{\alpha / \rho}$.

However, in the situation where $2\beta-\alpha<0$, 
we may increase $v$ from $0$ only up to $\tilde{v}$ (where $r=0$) 
in order to satisfy \eqref{decay}, where $\tilde{v}$ is defined by 
\begin{equation}
\rho \tilde{v}^2 = 2[\beta-\alpha + \sqrt{2\alpha(\alpha-\beta)}]
 < \alpha = \rho v_L^2.
\end{equation}
Clearly  the limiting speed is then $\hat{v}=\tilde{v}$.

Conducting a similar analysis for the upper half-space, we conclude
that the limiting speed is
\begin{equation}
\hat{v} = 
\left\{
\begin{array}{lll}
& \text{min}\{v_L, v^*_L\},  & \text{when} \quad
    2\beta > \alpha,  2\beta^* > \alpha^*, 
 \\
& \text{min}\{\tilde{v}, v^*_L\},  & \text{when} \quad
    2\beta < \alpha,  2\beta^* > \alpha^*,
\\ 
& \text{min}\{v_L, \tilde{v}^*\}, \quad & \text{when} \quad
    2\beta > \alpha,  2\beta^* < \alpha^*,
 \\ 
& \text{min}\{\tilde{v}, \tilde{v}^*\}, & \text{when} \quad
    2\beta < \alpha,  2\beta^* < \alpha^*.
\end{array}
\right.
\end{equation}

Note that the analysis conducted in this subsection (defining the 
subsonic interval) and in the following one (defining the conditions
of existence of a Stoneley wave) rely directly upon the methods 
developed by Chadwick \cite{Chad95} for compressible materials,
with the modifications required to accommodate the Bell constraint.

\subsection{Existence and uniqueness of interfacial (Stoneley) waves; 
comparison with surface (Rayleigh) waves}

Here we derive the conditions of existence for a Stoneley wave at the
interface of two deformed Bell-constrained half-spaces,
using a ``matrix reformulation'' of the secular equation 
\eqref{secular}, a method suggested by Chadwick \cite{Chad95} and
based on the surface impedance method of Barnett \textit{et al.}
\cite{BLGM85}.
Indeed, introducing the following symmetric $2 \times 2$ matrices
\begin{equation}
\mathbf{M}(v) = 
 \gamma \begin{bmatrix}
 \sqrt{\frac{\alpha}{\gamma}}\eta r 
          & 1-\sqrt{\frac{\alpha}{\gamma}}\eta
\\
 1-\sqrt{\frac{\alpha}{\gamma}}\eta & \sqrt{\frac{\alpha}{\gamma}} r
        \end{bmatrix},
\quad
\mathbf{M}^*(v) = 
 \gamma^* \begin{bmatrix}
 \sqrt{\frac{\alpha^*}{\gamma^*}}\eta^* r^* 
       & \sqrt{\frac{\alpha^*}{\gamma^*}}\eta^* -1
\\
 \sqrt{\frac{\alpha^*}{\gamma^*}}\eta^* - 1
       & \sqrt{\frac{\alpha^*}{\gamma^*}} r^*
        \end{bmatrix},
\end{equation}
we find that the secular equation \eqref{secular} corresponds to 
\begin{equation} \label{i=det(N)}
i(v) =  \text{det } \mathbf{N}(v) = 0, \quad \text{where} \quad
\mathbf{N}(v) = \mathbf{M}(v) + \mathbf{M}^*(v).
\end{equation}
From then on, it is an easy matter to transpose Chadwick's results 
\cite{Chad95} for incompressible materials to Bell materials, and to 
show \textit{inter alia} that the eigenvalues of 
$\mathbf{N}(v)$ decrease monotonically as $v$ increases in $I$.
Then the following results apply (see also Barnett \textit{et al.} 
\cite{BLGM85} for Stoneley waves in linear anisotropic elasticity):
when an interfacial wave exists, it is unique; 
it propagates at a speed which is greater than the speed of the 
Raleigh wave associated with either each half-space; 
it exists if and only if 
\begin{equation} \label{conditions}
i(0)>0, \quad i(\hat{v}) <0.
\end{equation}
In the ($\lambda_1, \lambda_2, \lambda_3,
           \lambda^*_1, \lambda^*_2, \lambda^*_3$)-space,
the curve $i(\hat{v}) = 0$ is called the \textit{limiting equation} 
and the curve $i(0)=0$ is the \textit{neutral equation}.
Dowaikh and Ogden \cite{DoOg91} refer to this latter equation as the
``exclusion equation'' because it rules out the possibility of 
incremental inhomogeneous static deformations in homogeneously 
deformed half-spaces. 
Biot \cite{Biot63} called it the ``characteristic equation for 
instability'', because it ``corresponds to the spontaneous appearance
of sinusoidal deformations at the interface.''
We obtain this equation for Bell materials by taking 
$\eta = \eta^* =1$ in the secular equation \eqref{secular}:
\begin{equation} \label{neutralCondition}
\sqrt{2\gamma(\beta+\alpha)} + \sqrt{2\gamma^*(\beta^*+\alpha^*)} 
\pm (\sqrt{\alpha \gamma}
 - \sqrt{\alpha^* \gamma^*} +\gamma^* - \gamma ) = 0.
\end{equation} 

%
\section{UNIFORM PREDEFORMATION AND \\ SPE\-CI\-FIC 
STRAIN ENERGY FUNC\-TIONS}
%

In the preceding section, we covered some ground on the propagation of 
Stoneley waves in deformed Bell materials.
We obtained results for general prestrains and unrestricted strain 
energy density functions.
We now turn our attention to special configurations.
First, we assume that the infinite body made of the two bonded 
Bell-constrained semi-infinite bodies is predeformed uniformly in 
the whole space, so that the principal stretches are the same for the
lower and upper half-spaces:
\begin{equation} 
\lambda_1^* = \lambda_1, \quad 
\lambda_2^* = \lambda_2, \quad 
\lambda_3^* = \lambda_3.
\end{equation}
This deformation is possible with the application of the following
loads,
\begin{align} 
& P_\Gamma =  (\lambda_2 -  \lambda_\Gamma)
         (-\omega_0 + \lambda_\Gamma \lambda_2 \omega_2)/ \lambda_2,
\notag \\
& P^*_\Gamma =  (\lambda_2 -  \lambda_\Gamma) 
   (-\omega^*_0 +\lambda_\Gamma\lambda_2 \omega^*_2)/\lambda_2,
\notag \\
& (\Gamma =1,3).
\end{align}

Next, we specialize the analysis to specific forms of the strain
energy functions and make the connection with historical results
obtained for incompressible materials.

\subsection{Simple hyperelastic Bell materials}

Here we consider that the lower and upper half-spaces are made of
``Simple hyperelastic Bell'' materials, for which the strain energy
functions are of the form \cite{BeHa92a},
\begin{multline}
W_\text{SHB}=\mathcal{C}_1
   (3 - \lambda_1\lambda_2 -\lambda_2\lambda_3 - \lambda_3\lambda_1) 
       + \mathcal{C}_2 (1 - \lambda_1\lambda_2\lambda_3),
\\
W^*_\text{SHB}=\mathcal{C}^*_1
   (3 - \lambda_1\lambda_2 -\lambda_2\lambda_3 - \lambda_3\lambda_3) 
       + \mathcal{C}^*_2 (1 - \lambda_1\lambda_2\lambda_3). 
\end{multline}
The material constants $\mathcal{C}_i$, $\mathcal{C}^*_i$
satisfy, according to the $A$-inequalities \eqref{A-inequalities},
\begin{equation}
\mathcal{C}_1>0, \quad \mathcal{C}_2 \ge 0, \quad
\mathcal{C}^*_1>0, \quad \mathcal{C}^*_2 \ge 0.
\end{equation}
Now, the expressions \eqref{AlphaBetaGamma} and \eqref{r-eta}$_2$
for $\alpha$, $\beta$, $\gamma$, and $r$  
(and for $\alpha^*$, $\beta^*$, $\gamma^*$, and $r^*$) reduce to 
\begin{equation}
\gamma = (\mathcal{C}_1\lambda_3^{-1} + \mathcal{C}_2)
           \lambda_2/(\lambda_1+\lambda_2)>0,
\quad
\alpha = \beta = \lambda_1^2\lambda_2^{-2}\gamma>0,
\quad
r=\eta + 1.
\end{equation}
Because here $2\beta - \alpha = \alpha >0$, we deduce from the 
discussion undertaken in \S\ref{The-subsonic-interval} that the 
limiting speed for two bonded simple hyperelastic Bell materials is
\begin{equation}
\hat{v} = \text{min}\{\sqrt{\alpha/\gamma},\sqrt{\alpha^*/\gamma^*}\}.
\end{equation}
For $v \in I=[0,\hat{v}]$, the secular equation \eqref{secular} 
reduces to
\begin{align} \label{secularSHB}
\lambda_1^{-2}\lambda_2^2 i(v) & \equiv  
\gamma^2[\eta^3 + \eta^2 + (1+2\lambda_1^{-1}\lambda_2)\eta 
          - \lambda_1^{-2}\lambda_2^2]
 \nonumber
\\
 &  \quad +
\gamma^{*2}[\eta^{*3} + \eta^{*2} + (1+2\lambda_1^{-1}\lambda_2)\eta^* 
          - \lambda_1^{-2}\lambda_2^2]
 \\
 & \quad  \quad + \gamma \gamma^*
[(\eta + \eta^*)(\eta + 1)(\eta^* + 1)
 +2(\lambda_1^{-1}\lambda_2 - \eta)(\lambda_1^{-1}\lambda_2-\eta^*)]
 \nonumber \\
& =  0. \nonumber
\end{align}

The conditions of existence of a Stoneley wave are \eqref{conditions},
with the above simplifications.
Explicitly, the neutral condition $i(0)>0$ factorizes to
\begin{equation} \label{neutralSHB1}
[(3\gamma^* + \gamma)\lambda_1 - (\gamma^* - \gamma)\lambda_2]
 [(3\gamma + \gamma^*)\lambda_1 - (\gamma - \gamma^*)\lambda_2]>0.
\end{equation}
The sign of each factor depends on the sign of $\gamma - \gamma^*$.
Introducing the quantity $\epsilon$ defined by
\begin{equation}
\epsilon = 
\left\{
\begin{array}{lll}
& \gamma/\gamma^* = 
   (\mathcal{C}_1 + \mathcal{C}_2\lambda_3)/
	  (\mathcal{C}^*_1 + \mathcal{C}^*_2\lambda_3)
 & \text{when} \quad \gamma < \gamma^*, 
 \\ 
& \gamma^*/\gamma = 
   (\mathcal{C}^*_1 + \mathcal{C}^*_2\lambda_3)/
	  (\mathcal{C}_1 + \mathcal{C}_2\lambda_3)  
 & \text{when} \quad \gamma^* < \gamma, 
\end{array}
\right.
\end{equation}
we write the neutral condition \eqref{neutralSHB1} as
\begin{equation} \label{neutralSHB2}
(3+\epsilon)\lambda_1 - (1-\epsilon)\lambda_2>0.
\end{equation}
Note that the neutral condition \eqref{neutralSHB2} gives an 
implicit relationship between the stretch ratios $\lambda_1$ and 
$\lambda_2$ because $\lambda_3$ appearing in $\epsilon$ is linked to
$\lambda_1$ and $\lambda_2$ through the Bell constraint:
$\lambda_3 = 3 - \lambda_1 - \lambda_2$ (explicitly, the relationship
between  $\lambda_1$ and $\lambda_2$ is quadratic.)
However, for the plane strain $\lambda_3=1$ (and 
$\lambda_2=2-\lambda_1$), $\epsilon$ is independent of the stretch 
ratios and the critical stretch $(\lambda_1)_\text{cr}$, 
at which the neutral equation is reached, is
\begin{equation} \label{planeSHB}
(\lambda_1)_\text{cr} = \textstyle{\frac{1}{2}} (1-\epsilon).
\end{equation}
Note also that when one half-space is absent
($\gamma = 0$ or $\gamma^*=0$) then $\epsilon = 0$ in 
\eqref{neutralSHB2} and we recover the relative universal surface 
stability condition $3\lambda_1 - \lambda_2>0$ for the remaining 
half-space \cite{Dest03}.

Finally we express the limiting condition $i(\hat{v})>0$ in the case
where 
$v^*_L = \sqrt{\alpha^*/\rho^*} < v_L = \sqrt{\alpha/\rho}$.
Then we have
\begin{equation}
\hat{v}=v^*_L = \sqrt{\frac{\alpha^*}{\rho^*}}, 
\quad \eta^*(\hat{v}) = \eta^*_L = 0, \quad
\eta(\hat{v}) = \eta_L = \sqrt{\frac{\alpha-\rho v_L^{*2}}{\alpha}}
 = \sqrt{1 - \frac{\rho \alpha^*}{\rho^* \alpha}},
\end{equation}
and the limiting condition $i(\hat{v})>0$ is
\begin{equation} \label{limitingSimple}
\eta_L(\eta_L+1)(\eta_L+1+\gamma^*/\gamma)
 - [\eta_L - \lambda_1^{-1}\lambda_2(1-\gamma^*/\gamma)]^2<0.
\end{equation}
In the case where 
$v_L = \sqrt{\alpha/\rho} < v^*_L = \sqrt{\alpha^*/\rho^*}$,
the starred and unstarred quantities above are interverted.

In his seminal paper, Stoneley \cite{Ston24} showed, by means of 
numerical examples, that the existence of `waves at the surface of 
separation of two solids' depended heavily on the material properties 
of each half-space.
We treat two numerical examples below, choosing 
$\rho$, $\alpha$, $\gamma$, $\rho^*$, $\alpha^*$, $\gamma^*$, such
that a connection is made with his results.

\vspace{12pt}

\noindent
\textit{Example 1: $\alpha^*/\rho^* = \alpha/\rho$, 
with $\rho = 8.2$, $\rho^* = 3.2$}.

In this case, we have
\begin{equation} \label{example1}
\epsilon = \frac{\gamma^*}{\gamma} = \frac{\alpha^*}{\alpha}
   = \frac{\rho^*}{\rho} = \frac{3.2}{8.2} = 0.39024.
\end{equation}
Important simplifications occur in this special case, as
\begin{align}
& \eta = \eta^* = \sqrt{1 - \frac{\rho v^2}{\alpha}} = 
  \sqrt{1 - \frac{\rho^* v^2}{\alpha^*}}, \notag \\
& \hat{v} = v_L = v_L^* = \sqrt{\frac{\alpha}{\rho}} 
                    =  \sqrt{\frac{\alpha^*}{\rho^*}}, \notag \\
& \eta_L = \eta_L^* =0.
\end{align}
Hence, the limiting condition $i(\hat{v})>0$ is automatically 
satisfied, because $\eta_L=0$ in \eqref{limitingSimple}.
The only condition of existence of a Stoneley wave is the neutral 
condition \eqref{neutralSHB2}, that is
\begin{equation} \label{neutralExample1}
\lambda_1 - 0.17986 \lambda_2 >0.
\end{equation}
Finally, the secular equation \eqref{secularSHB} reduces further to
\begin{equation} \label{secularExample1}
(1+\epsilon)^2 
  [\eta^3 + \eta^2 + (1+2\lambda_1^{-1}\lambda_2)\eta 
          - \lambda_1^{-2}\lambda_2^2]
 +
4\epsilon(\eta - \lambda_1^{-1}\lambda_2)^2=0,
\end{equation}
with $\epsilon$ given in \eqref{example1}.
In particular, when the materials are undeformed in the static state
($\lambda_1 = \lambda_2 = \lambda_3 = 1$), the constraint of Bell 
coincides with the constraint of incompressibility \cite{BeHa92a}, 
and we recover Stoneley's result \cite{Ston24}: 
$v = 0.99287 \hat{v}$,
for linear isotropic incompressible materials with material parameters
matching those of this example.

Figure 1(a) shows the intersection of the triangle 
$\lambda_1+\lambda_2+\lambda_3=3$ of possible values for the stretch
ratios in Bell materials with the neutral condition 
\eqref{neutralExample1} for the combination of simple hyperelastic 
Bell materials in this example;
the visible part of the triangle represents the configurations where 
a Stoneley wave exists, and where the interface is stable with respect
to incremental deformations.

Figure 1(b) illustrates the influence of the loads on the Stoneley 
wave speed (thick curve) in the case of plane strain $\lambda_3=1$.
Then $\lambda_2 = 2 - \lambda_1$, and the critical load obtained from 
\eqref{neutralExample1} is $(\lambda_1)_\text{cr} = 0.30488$.
The speed, scaled with respect to the limiting wave speed 
$\hat{v} = \sqrt{\alpha/\rho} = \sqrt{\alpha^*/\rho^*}$,
is the coordinate on the vertical axis; the stretch ratio $\lambda_1$
is the coordinate on the horizontal axis.
In tension ($1\le \lambda_1 < 2$) the Stoneley wave propagates at a
speed which is within less than $1\%$ of the limiting speed;
under increasing compression 
($1\ge \lambda_1 > (\lambda_1)_\text{cr}$), the speed drops rapidly 
to zero.
The Figure also shows the scaled Rayleigh wave speed associated with
either half-space \cite{Dest03}, computed by taking $\epsilon = 0$ in
\eqref{secularExample1} (thin curve, always below the Stoneley wave
speed curve as expected).

\begin{figure}
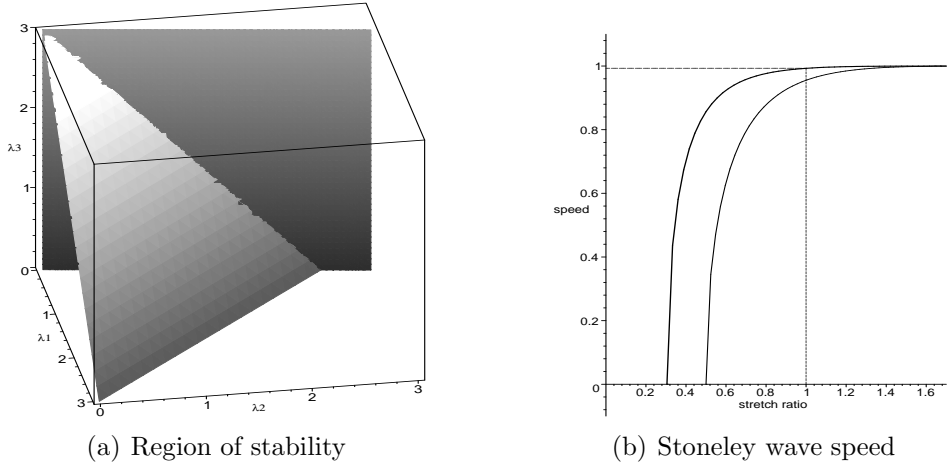

\centering
\mbox{\subfigure[Region of stability]{\epsfig{figure=example1a.eps,
width=.40\textwidth, height=.40\textwidth}}
 \quad \quad \quad \quad 
      \subfigure[Stoneley wave speed]{\epsfig{figure=example1b.eps,
width=.37\textwidth, height=.37\textwidth}}}
\caption{Simple hyperelastic Bell materials 
($\alpha^*/\rho^* = \alpha/\rho$, $\rho = 8.2$, $\rho^* = 3.2$).}
\end{figure}

\vspace{12pt}

\noindent
\textit{Example 2: $\alpha^*/\rho^* 
 =\textstyle{\frac{3}{2}}\alpha/\rho$, with $\rho = 2$, $\rho^* = 1$}.

In this case, we have
\begin{equation} \label{example2}
\epsilon = \frac{\gamma^*}{\gamma} = \frac{\alpha^*}{\alpha}
   = \frac{3\rho^*}{2\rho} = \frac{3}{4}.
\end{equation}
The limiting speed is
\begin{equation}
\hat{v} = v_L  = \sqrt{\frac{\alpha}{\rho}} <  
  v_L^*= \sqrt{\frac{\alpha^*}{\rho^*}}= \sqrt{\frac{3\alpha}{2\rho}}, 
\end{equation}
and the corresponding $\eta$'s are 
\begin{equation}
\eta_L = 0, \quad 
\eta_L^* = \sqrt{1 - \frac{\rho^*v_L^2}{\alpha^*}} 
         = \sqrt{1 - \frac{\rho^*\alpha}{\rho \alpha^*}} 
 =  \frac{1}{\sqrt{3}}.
\end{equation}

The limiting condition \eqref{limitingSimple}  (with the
starred and unstarred quantities interverted) is now satisfied for
$\lambda_1$, $\lambda_2$ such that 
\begin{equation} \label{limitingExample2}
(\sqrt{10+8\sqrt{3}} - \sqrt{3})\lambda_1 - \lambda_2 <0,
\quad
\text{or}
\quad
\lambda_1 -  0.31723\lambda_2 <0.
\end{equation}
Here we see that no interface wave may propagate when the materials
are undeformed in the static state 
($\lambda_1 = \lambda_2 = \lambda_3 = 1$), as noted by Stoneley
\cite{Ston24} for linear isotropic incompressible materials with 
material parameters matching those of this example.

The other condition of existence of a Stoneley wave is the neutral 
condition \eqref{neutralSHB2}, that is here
\begin{equation} \label{neutralExample2}
15 \lambda_1 - \lambda_2 >0.
\end{equation}

Figure 2(a) shows the intersection of the triangle 
$\lambda_1+\lambda_2+\lambda_3=3$ of possible values for the stretch
ratios in Bell materials with the limiting condition 
\eqref{limitingExample2} and with the neutral condition 
\eqref{neutralExample2} for the combination of simple hyperelastic 
Bell materials in this example;
the plane of the Figure coincides with the plane of the triangle.
We see that the range of possible stretch ratios is far smaller here 
than in the previous example. 
For instance, in the case of plane strain $\lambda_3=1$, the stretch
ratio $\lambda_1$ must belong to the (compressive) interval
$]0.125, 0.48167[$.
Note that in this plane strain case, the Stoneley wave exists in 
a range where Rayleigh waves do not exist (in \cite{Dest03} it is 
proved that the domain of existence of Rayleigh waves in plane strain
is: $\lambda_1 \in ]0.5, 2[$.)

Figure 2(b) illustrates the influence of the loads on the Stoneley 
wave speed (thick curve) in the case of plane strain $\lambda_3=1$.
The stretch ratio $\lambda_1$ is the coordinate on the horizontal 
axis.
The speed, scaled with respect to the limiting wave speed 
$\hat{v} = \sqrt{\alpha/\rho}$, is the coordinate on the vertical axis;
it is computed by solving numerically \eqref{secularSHB} for 
$y = v/\hat{v}$ with $\eta = \sqrt{1-y^2}$, 
$\eta^* = \sqrt{1- \textstyle{\frac{2}{3}}y^2}$, and 
$\gamma^* = \textstyle{\frac{3}{4}}\gamma$.
Also represented are the Rayleigh wave speeds (thin curves) associated 
with each half-space, computed by solving numerically 
\eqref{secularSHB} for $y = v/\hat{v}$ with $\eta = \sqrt{1-y^2}$ 
and $\gamma^*= 0$ (lower thin curve, with $y=1$ as an horizontal 
asymptote) or $\eta^* = \sqrt{1- \textstyle{\frac{2}{3}}y^2}$ and 
$\gamma = 0$ (upper thin curve, with 
$y=\sqrt{\textstyle{\frac{3}{2}}}$ as an horizontal asymptote).

\begin{figure}
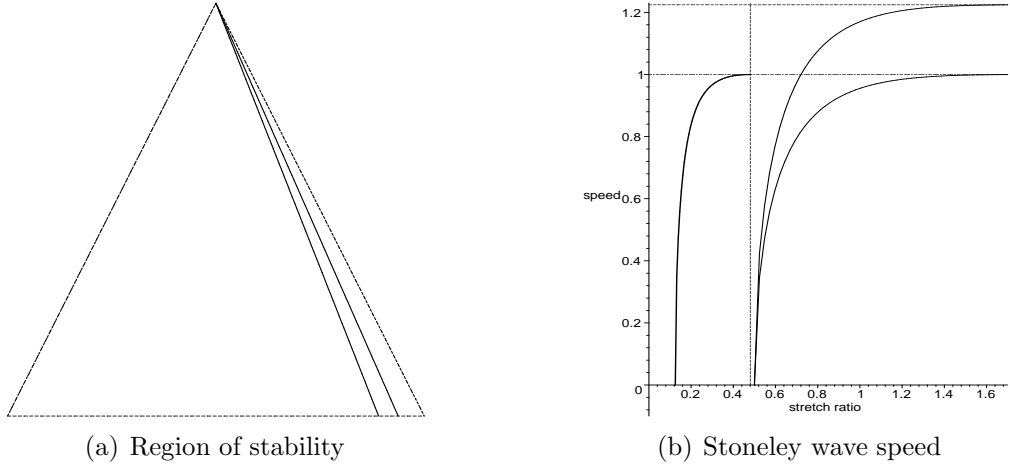

\centering
\mbox{\subfigure[Region of stability]{\epsfig{figure=example2a.eps,
height=.40\textwidth, width=.40\textwidth}}
 \quad \quad \quad \quad \quad
      \subfigure[Stoneley wave speed]{\epsfig{figure=example2b.eps,
height=.40\textwidth, width=.40\textwidth}}}
\caption{Simple hyperelastic Bell materials 
($\alpha^*/\rho^* = \textstyle{\frac{3}{2}}\alpha/\rho$, $\rho = 2$, 
$\rho^* = 1$).}
\end{figure}

\subsection{Bell's empirical model}

For \textit{Bell's empirical model materials} \cite{BeHa92a}, 
the strain energy functions of each half-space are
\begin{align} \label{BellEmpirical}
& W_\text{BEM} = \textstyle{\frac{2}{3}} \nu_0 
 [2(3-\lambda_1\lambda_2 + \lambda_2\lambda_3 + 
                  \lambda_3\lambda_1)]^{\textstyle{\frac{3}{4}}},
\notag \\
& W^*_\text{BEM} = \textstyle{\frac{2}{3}} \nu^*_0 
 [2(3-\lambda_1\lambda_2 + \lambda_2\lambda_3 + 
                  \lambda_3\lambda_1)]^{\textstyle{\frac{3}{4}}},
\end{align}
where $\nu_0$, $\nu^*_0$ are positive constants.
For these models, the material response functions $\omega_0$, 
$\omega_2$, $\omega^*_0$ and $\omega^*_2$ provided by 
\eqref{A-inequalities} are 
\begin{equation}
\omega_0 = \omega^*_0 = 0,
\quad
\omega_2 =i_3^{-1} \nu_0 [2(3-i_2)]^{\textstyle{-\frac{1}{4}}}, 
\quad
 \omega_2^* = \epsilon \omega_2, 
\quad
\epsilon = \nu_0^*/\nu_0,
\end{equation}
with $i_2$, $i_3$, given by \eqref{invariants-static}$_{3,4}$.
Then $\alpha$, $\beta$, $\gamma$, $\alpha^*$, $\beta^*$, $\gamma^*$
are computed as: 
\begin{align} \label{coeffsBEM}
&\alpha = \frac{\lambda_1^3}{\lambda_1 + \lambda_2}\omega_2
 = \frac{\lambda_1^2}{\lambda_2^2}\gamma,
\quad
\beta = \frac{\lambda_1^2}{2}
  \Big{[}2 - \frac{(\lambda_1 - \lambda_2)^2}{4(3-i_2)}\Big{]} \omega_2
    - \frac{\lambda_1^2 \lambda_2}{\lambda_1+\lambda_2} \omega_2,
\\
&\alpha^* = \epsilon \alpha,
\quad
\beta^* = \epsilon \beta,
\quad
\gamma^* = \epsilon \gamma.
\end{align}

Although the strain energy function of a ``Bell's empirical model''
material depends only upon one material constant ($\nu_0$), 
as opposed to two ($\mathcal{C}_1$ and $\mathcal{C}_2$) for a 
``simple hyperelastic Bell'' material, 
the full analysis of the Stoneley wave existence is too cumbersome 
and lengthy to be followed here.
Indeed, we saw in \S\ref{The-subsonic-interval} that the definition
of the limiting speed depends on the sign of $2\beta - \alpha$
which, for these materials, is equal to
\begin{equation}
2\beta - \alpha = \omega_2 \lambda_1^2
  [ 3 \lambda_1^3 - (12-5\lambda_2)\lambda_1(\lambda_1+ \lambda_2)
         - \lambda_2^3]/[4(3-i_2)(\lambda_1+ \lambda_2)].
\end{equation}
The sign of this quantity (and hence the definition of $\hat{v}$)
depends heavily on the values of the stretches $\lambda_1$ and 
$\lambda_2$ (For instance in the equibiaxial case
 $\lambda_1 = \lambda_3$, we have 
$2\beta - \alpha = \omega_2 \lambda_1^2
  (7 \lambda_1 - 9)/[4(3-\lambda_1)]$, which changes sign at 
$\lambda_1 = \textstyle{\frac{9}{7}}$.)
Consequently, the limiting condition is difficult to obtain in 
general.
Moreover, the connection with Stoneley's results \cite{Ston24}
cannot be made because the parameters $\omega_2$, $\omega_2^*$ 
are singular when the material is isotropic (then $i_2=3$),
where a different approach must be adopted \cite{BeHa92a}.

Nevertheless, the neutral equation, which is 
independent of $\hat{v}$ and takes place away from isotropy,
can be obtained and compared with Biot's ``stability equation'' for
rubber-like incompressible materials \cite{Biot63}.
We deduce it by specializing  \eqref{neutralCondition} to the values
\eqref{coeffsBEM} of $\alpha$, $\beta$, $\gamma$, $\alpha^*$, 
$\beta^*$, $\gamma^*$, as
\begin{equation} \label{neutralBEM}
\frac{4\lambda_1}{\lambda_1 + \lambda_2}
 - \frac{(\lambda_1 - \lambda_2)^2}{4(3-i_2)}
- \Big{[}\frac{1-\epsilon}{1+\epsilon} \Big{]}^2
    \frac{(\lambda_1 - \lambda_2)^2}{\lambda_1(\lambda_1 + \lambda_2)}
 = 0.
\end{equation}
In the case of plane strain $\lambda_3=1$, the equation reduces to
\begin{equation} \label{neutralBEMplane}
(2\lambda_1 -1 )\lambda_1 - 2
 \Big{[}\frac{1-\epsilon}{1+\epsilon} \Big{]}^2
    (\lambda_1 - 1)^2
 = 0.
\end{equation}
Note that when $\epsilon = 0$ in \eqref{neutralBEM}, 
only the lower half-space subsists and we recover the surface 
bifurcation criterion for Bell's empirical materials in compression 
\cite{BePa98},
\begin{equation} 
3 - \frac{(\lambda_1 - \lambda_2)^2}{4(3-i_2)}
-\lambda_1^{-1} \lambda_2 = 0.
\end{equation}

Figure 3 (where the plane of the Figure coincides with the plane of 
the triangle) shows the intersection between the triangle of possible
stretch ratios for Bell materials and the neutral curve in the 
cases where $\epsilon = 0.0$ \cite{Dest03} (thickest curve),
$\epsilon = 0.2$, $\epsilon = 0.4$, $\epsilon = 0.6$ (thinnest curve),
and $\epsilon = 0.8$ (dotted curve).
\begin{figure}
\begin{centering}
\epsfig{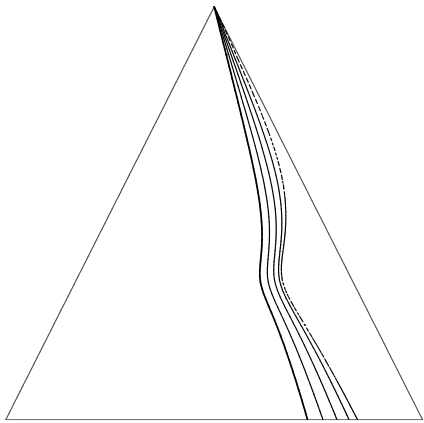}
\caption{Neutral curves for `Bell empirical model'
materials.}
\end{centering}
\end{figure}

\subsection{Comparison with neo-Hookean incompressible materials}

Biot \cite{Biot63} investigated the surface instability of 
incompressible materials under finite compression, and obtained
the neutral condition, which he called the ``characteristic 
equation for interfacial instability.''
He then discussed the equation in the particular case of a neo-Hookean
strain energy function,
\begin{equation} 
 W_\text{nH} = \mu_0
                  (\lambda_1^2 + \lambda_2^2 + \lambda_3^2 - 3),
\end{equation}
and expressed it, in the case of plane strain $\lambda_3=1$, as:
\begin{equation}
\lambda_1^2\Big{[}\frac{1+\lambda_1^2}{1-\lambda_1^2}\Big{]}^2 
= \Big{[}\frac{1-\epsilon}{1+\epsilon}\Big{]}^2, \quad
\epsilon = \frac{\mu_0}{\mu_0^*}.
\end{equation}
He did not mention that the result was also valid for the neo-Hookean
strain energy function,
\begin{equation} 
 W_\text{MR} = \mathcal{D}_1
                  (\lambda_1^2 + \lambda_2^2 + \lambda_3^2 - 3) 
  + \mathcal{D}_2(\lambda_1^2\lambda_2^2 + \lambda_2^2\lambda_3^2 
       + \lambda_3^2\lambda_1^2 - 3),
\end{equation}
by replacing the rigidity ratio $\epsilon$ with 
\begin{equation}
\epsilon =
 \frac{\mathcal{D}^*_1-\mathcal{D}^*_2}{\mathcal{D}_1-\mathcal{D}_2}.
\end{equation}

Table 1 shows the numerical values of the critical stretches  
for the classes of Bell's empirical model (2nd column), 
of neo-Hookean and Mooney-Rivlin incompressible materials 
(3rd column), and of simple hyperelastic Bell materials (4th column) 
in the case of plane strain, for different values of the 
rigidity ratio $\epsilon$.
It appears that rubber can be compressed more than Bell's empirical 
model but less than simple hyperelastic Bell materials, 
before the neutral equation is satisfied.

\begin{center}
{\footnotesize
Table 1: Critical stretch ratios
for interface instability ($\lambda_3=1$)}
{\normalsize
\noindent
{\small
\begin{tabular}{l c c c}
\hline
\rule[-3mm]{0mm}{8mm} 
$\epsilon$ & Bell empirical  & rubber & simple Bell
\\
\hline
0.0 &  0.6667 & 0.5437 & 0.5000
\\
0.2 &  0.6105 & 0.4457 & 0.4000
\\
0.4 &  0.5625 & 0.3393 & 0.3000
\\
0.6 &  0.5266 & 0.2257 & 0.2000
\\
0.8 &  0.5060 & 0.1085 & 0.1000
\\ \hline
\end{tabular}
}}
\end{center}


\appendix 
 \renewcommand{\thesection}{\Alph{section}}

\section*{Appendix} 
\textit{Strong ellipticity conditions for the incremental equations
of motion in Bell materials.} 
\setcounter{section}{1} 
\setcounter{equation}{0} 

For an unconstrained hyperelastic material maintained in a static 
state of pure homogeneous deformation, the incremental equations of 
motion are
 \begin{equation} \label{eqnsOfMtn}
\mathcal{A}_{jilk} u_{k,jl} = \rho u_{i,tt},
\end{equation}
where $\mbox{\boldmath $\mathcal{A}$}$ is the fourth-order 
instantaneous linear elasticity tensor.
The nonzero components of $\mbox{\boldmath $\mathcal{A}$}$ are:
\begin{align}
& \mathcal{A}_{iijj} = \lambda_i \lambda_j W_{ij}, 
\nonumber \\
& \mathcal{A}_{ijij}
 =(\lambda_i W_i-\lambda_j W_j)\lambda_i^2/(\lambda_i^2-\lambda_j^2), 
 \\
& \mathcal{A}_{ijji} =  \mathcal{A}_{ijij} - \lambda_i W_i, 
\nonumber 
\end{align}
(no sum) when the underlying deformation has distinct principal 
stretch ratios $\lambda_1$, $\lambda_2$, $\lambda_3$.
The assumption of strong ellipticity for the equations of motion 
\eqref{eqnsOfMtn} imposes that
\begin{equation} \label{SEcompressible}
\mathcal{A}_{jilk} m_j n_i m_l n_k > 0,
\quad \text{for all nonzero} \quad \mathbf{m}, \textbf{n}.
\end{equation}

When the material is incompressible, $\text{det }\mathbf{V} = 1$, 
where $\mathbf{V}$ is the left stretch tensor, and an arbitrary 
pressure $P$ is introduced \cite{Chad97}.
Then the equations of motion are strongly elliptic for all values of 
$P$ when the inequalities \eqref{SEcompressible} hold subject to
\begin{equation}
\mathbf{m} \cdot \mathbf{n} = 0.
\end{equation}.
 
When the material is subject to the Bell constraint 
$\text{tr }\mathbf{V} = 3$, 
an arbitrary scalar $p$ is introduced \cite{BeHa92a}.
Then the equations of motion are strongly elliptic for all values of 
$p$ when the inequalities \eqref{SEcompressible} hold subject to
\begin{equation}
\mathbf{m} \cdot \mathbf{Vn} = 0.
\end{equation}
The unit vectors
\begin{equation}
\mathbf{m}
 = \lambda_1^{-\textstyle{\frac{1}{2}}} \cos\theta \mathbf{i} + 
     \lambda_2^{-\textstyle{\frac{1}{2}}} \sin \theta \mathbf{j},
\quad
\mathbf{n}
 = -\lambda_1^{-\textstyle{\frac{1}{2}}} \sin\theta \mathbf{i} + 
     \lambda_2^{-\textstyle{\frac{1}{2}}} \cos \theta \mathbf{j},
\quad
0 \le \theta \le 2 \pi,
\end{equation}
are two such vectors, and the strong ellipticity condition says that 
\begin{equation}
A \cos^4 \theta + 2B \sin^2 \theta \cos^2 \theta + C \sin^4 \theta >0,
\end{equation}
for all $\theta$, where $A$, $B$, $C$, are given by
\begin{align}
& A = \lambda_1^{-1}  \lambda_2^{-1} \mathcal{A}_{1212}, \quad
C = \lambda_1^{-1}  \lambda_2^{-1} \mathcal{A}_{2121},
\nonumber \\
& B 
 = (\lambda_1^{-2} \mathcal{A}_{1111} 
       + \lambda_2^{-2} \mathcal{A}_{2222})/2
   -  \lambda_1^{-1}  \lambda_2^{-1} (\mathcal{A}_{1122}
                                       + \mathcal{A}_{1221}).
\end{align}
Choosing $\theta = \arctan (A/C)^{-\textstyle{\frac{1}{4}}}$, 
we arrive at
$B + \sqrt{AC}>0$, which is equivalent to
\begin{equation}
\beta + \alpha >0,
\end{equation}
where
\begin{align}
& \alpha  =
    (W_1 - W_2) 
       \lambda_1^2 /[\lambda_2\lambda_3 (\lambda_1^2 - \lambda_2^2)], 
\notag \\
& \beta  = 
  (W_{11} + W_{22} - 2 W_{12}) \lambda_1/(2\lambda_2 \lambda_3)
    - \alpha \lambda_2/ \lambda_1.
\end{align} 

\end{document}